\documentclass[epjCONF]{svjour}
\usepackage{graphics}
\usepackage[varg]{txfonts} 
\usepackage[latin1]{inputenc}
\session-title{MENU2013}
\begin{document}
\title{New Analysis of Threshold Photoproduction Data from MAMI}
\author{C\'esar Fern\'andez-Ram\'{\i}rez\inst{1}\fnmsep\thanks{Current address: 
JLab Physics Analysis Center, Center for Theoretical and Computational Physics, 
Thomas Jefferson National Accelerator Facility, 12000 Jefferson Avenue, 
Newport News, VA 23606, USA, \email{cesar@jlab.org}}}
\institute{Grupo de F\'{\i}sica Nuclear, Departamento de F\'{\i}sica At\'omica, Molecular y Nuclear, 
Facultad de Ciencias F\'{\i}sicas, Universidad Complutense de Madrid, 
Avda. Complutense s/n, E-28040 Madrid, Spain}
\abstract{In this talk I will review the recently published results by 
the A2 and CB-TAPS Collaborations at MAMI on neutral pion photoproduction 
in the near-threshold region. The combined measurement of the differential cross section 
and the photon beam asymmetry with low statistical errors allowed for a precise determination 
of the energy dependence of the real parts of the S- and P-wave amplitudes for the first time, 
providing the most stringent test to date of the predictions of 
Chiral Perturbation Theory and its energy region of agreement with experiment.} 
\maketitle
\section{Introduction} \label{sec:intro}

The dynamical consequences of the spontaneous 
breaking of chiral symmetry in Quantum Chromodynamics (QCD) 
and the appearance of the $\pi$ meson as a pseudoscalar Nambu--Goldstone boson
are well known and many predictions are available in the literature \cite{book}. 
One of the most important is the softness of the
S-wave amplitude for the $\gamma N \rightarrow \pi^{0} N$ reaction in the near
threshold region which vanishes in the chiral limit \cite{CHPT}. 
On top of the softness of the S-wave, the P waves are expected to provide a large 
contribution due to the early appearance of the $\Delta$ resonance \cite{AB-Delta}
and also, because of the softness of the S wave, even D waves have an important
impact due to their interference with P waves \cite{FBD09a,FBD09b}.
Hence, the accurate extraction of the S and P waves from pion photoproduction data becomes
an important issue in the study of chiral symmetry breaking and hadron dynamics.

In order to test hadron dynamics in the low-energy regime and Chiral Perturbation Theory (CHPT)
predictions, A2 and CB-TAPS Collaborations  have run several experiments at MAMI (Mainz)
collecting accurate differential cross sections and photon beam asymmetries for the 
$\gamma p \rightarrow \pi^{0} p$ reaction in the near-threshold region \cite{Hornidge2013,CD12}. 
The high quality of these data allows to extract the S-wave and P-waves energy dependence and
to use the data to test current CHPT
and assess the energy range where the theory is accurate for this particular 
process \cite{FB13,BCHPT}.

\section{Partial Wave Analysis} \label{sec:results}
\subsection{Single-Energy Multipoles}
\begin{figure}
\begin{center}
\begin{tabular}{cc}
\rotatebox{-90}{\resizebox{0.34\columnwidth}{!}{\includegraphics{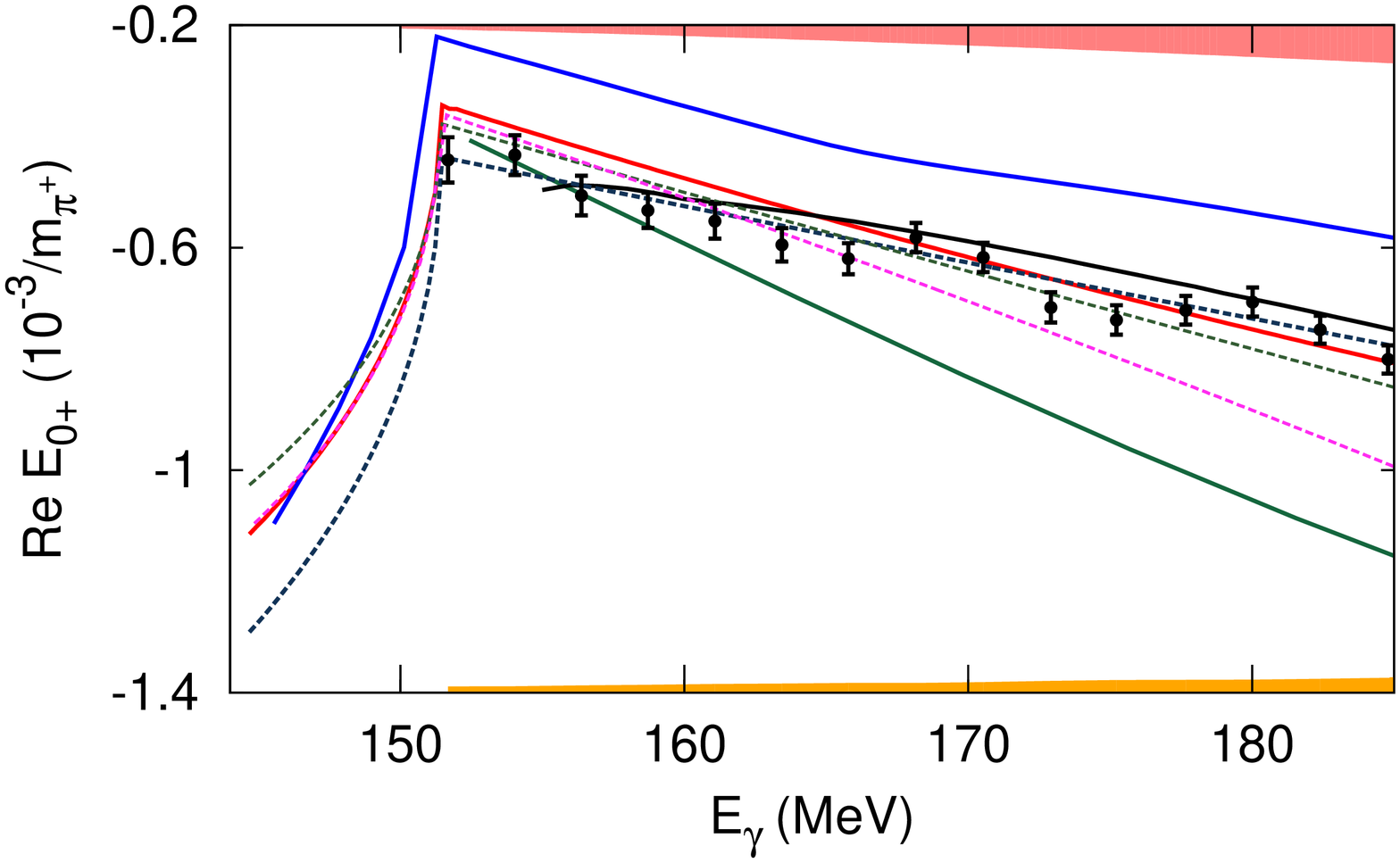} }} &
\rotatebox{-90}{\resizebox{0.34\columnwidth}{!}{\includegraphics{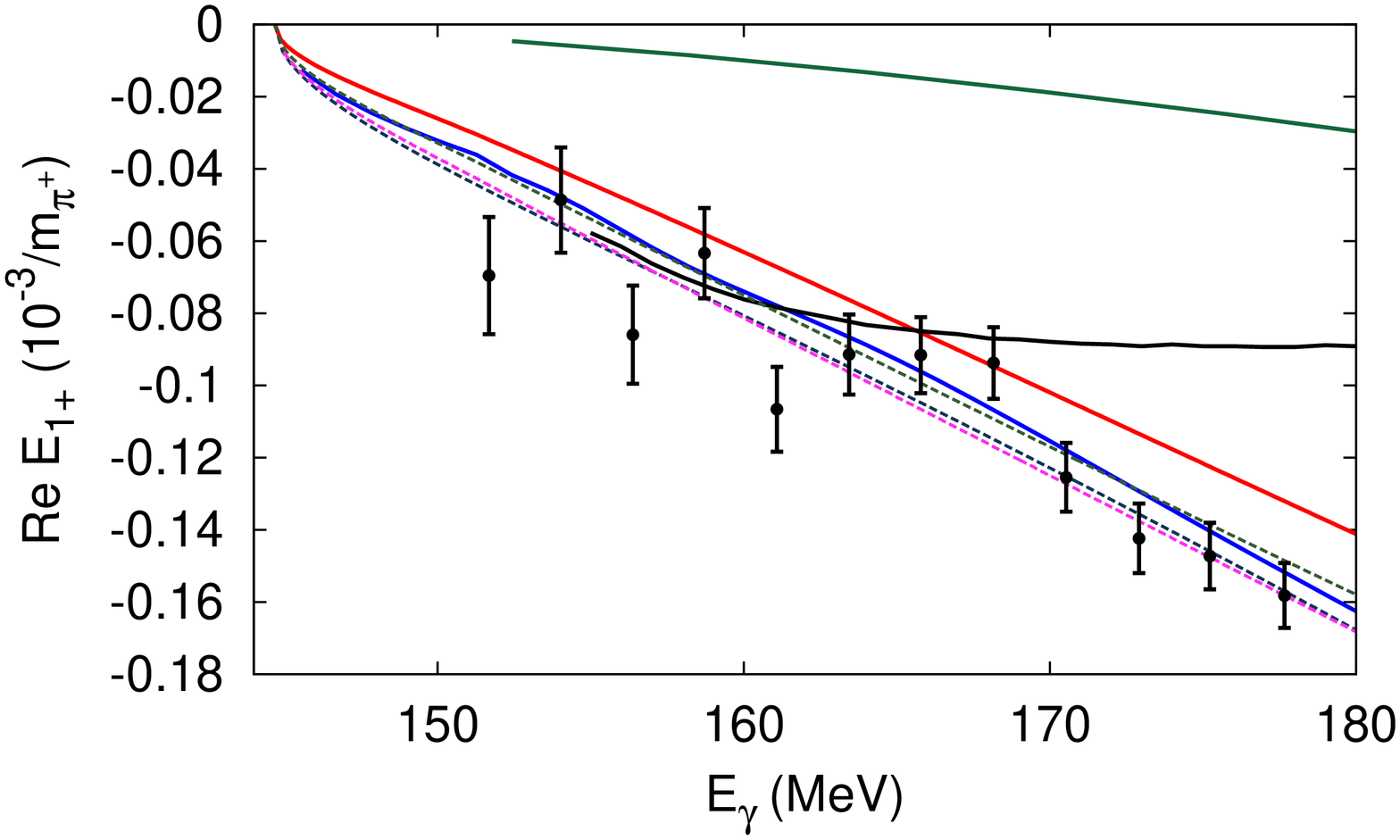} }} \\
\rotatebox{-90}{\resizebox{0.34\columnwidth}{!}{\includegraphics{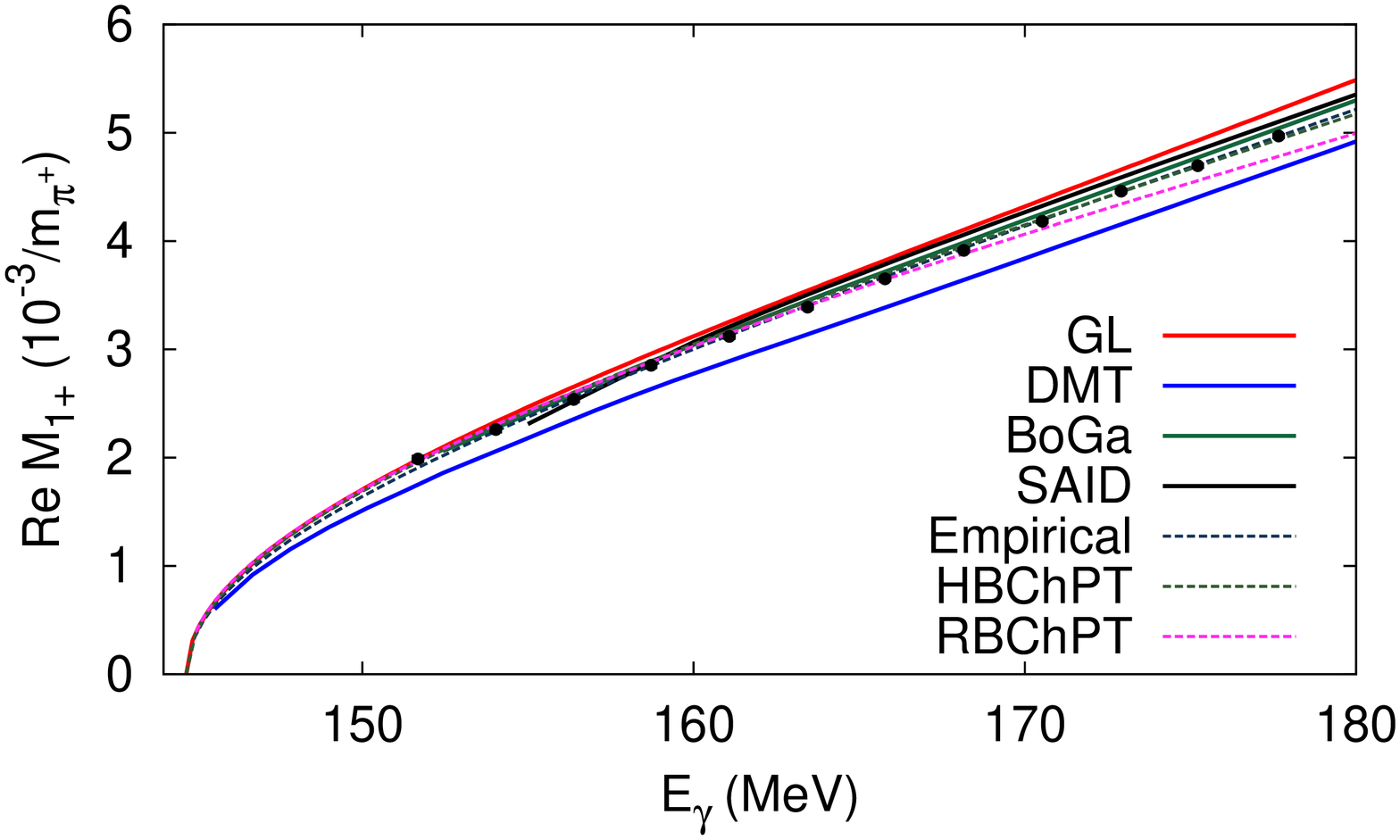} }} &
\rotatebox{-90}{\resizebox{0.34\columnwidth}{!}{\includegraphics{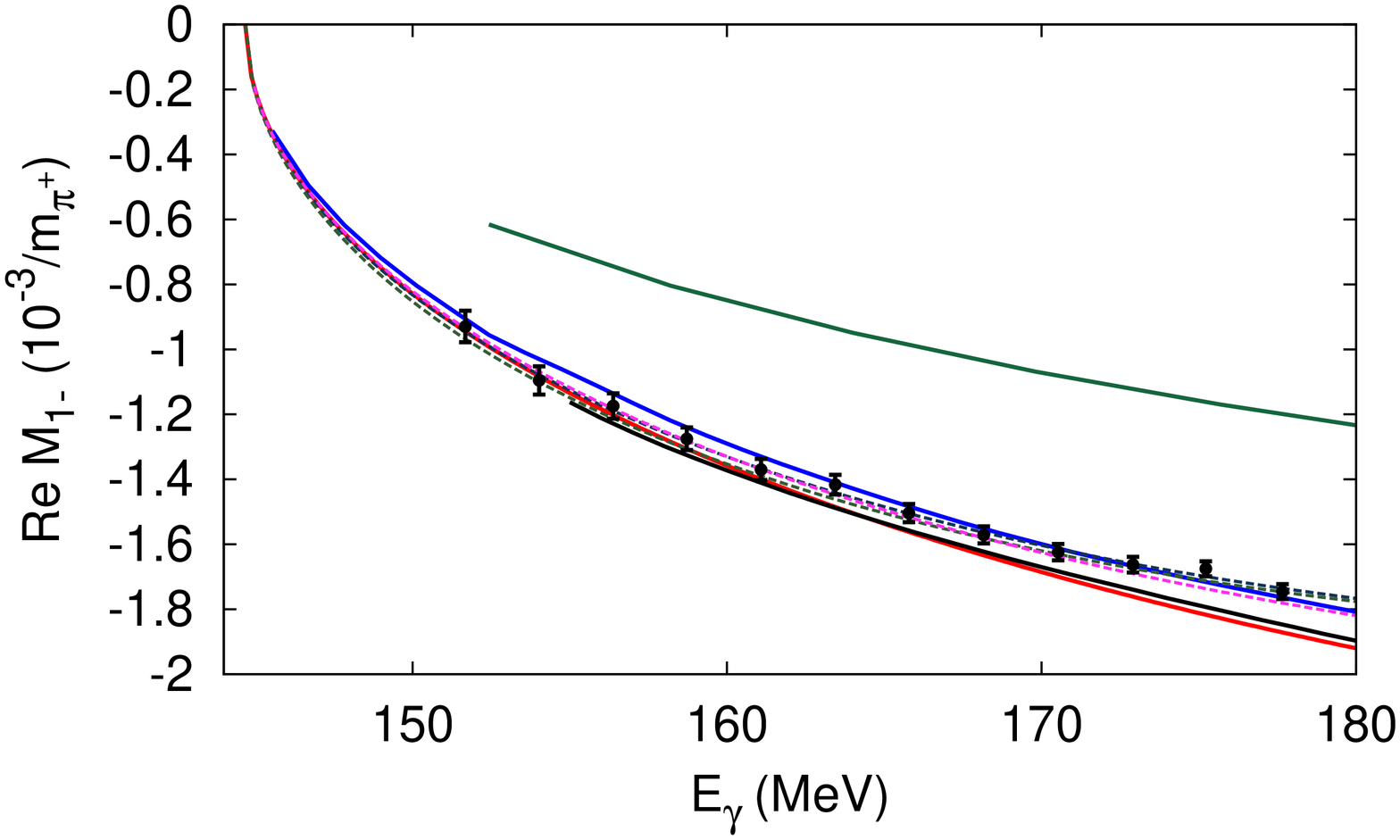} }} 
\end{tabular}
\end{center}
\caption{Extracted single-energy S and P partial waves 
($E_{0+}$, $E_{1+}$, $M_{1+}$ and $M_{1-}$: black dots) compared to fits to the experimental data 
(Empirical \cite{Hornidge2013} HBCHPT \cite{FB13} and RBCHPT \cite{BCHPT}: dashed lines) 
and to predictions:
Dispersive Effective Field Theory by Gasparyan and Lutz (GL) \cite{GL},
Dubna-Mainz-Taipei (DMT) \cite{DMT},
Bonn-Gatchina (BoGa) \cite{BoGa}, and
SAID (SN11 solution) \cite{SAID_SN11}: solid lines).
The red area at the top of the $E_{0+}$ figure represents 
the uncertainty in the single-energy multipoles due to D waves.}
\label{fig:experimental}       
\end{figure}

Data starting at $E_\gamma = 146$ MeV of photon energy in the laboratory frame
where collected approximately every $2.4$ MeV, obtaining simultaneously  for each energy
bin the differential cross section and the photon beam asymmetry. 
To extract the single-energy multipoles, each observable was fitted with the real part 
of $E_{0+}$, $E_{1+}$, $M_{1+}$, and $M_{1-}$ as free parameters
employing the algorithm described in  \cite{FMAA08}.
The  imaginary part of  $E_0+$ was set to the unitary value \cite{Hornidge2013,FB13} 
although the experiment is not sensitive to Im$E_{0+}$,
and the imaginary parts of $E_{1+}$, $M_{1+}$, and $M_{1-}$ were set to zero, 
which is an excellent approximation if we stay below 185 MeV where the $\Delta$
contribution to the imaginary part of $M_{1+}$ starts to play a role.
D waves were fixed to Born terms. 
The results are presented in Fig.  \ref{fig:experimental} together with several
fits and predictions which will be discussed in the next section.
The uncertainty due to our choice of D waves is presented
with a red band on the top of the Re$E_{0+}$ multipole plot.
D waves uncertainty makes no impact in the extraction of the P waves \cite{FBD09b}.
In this way, we have been able to extract 
single-energy S and P waves experimentally in a model independent way 
except for the uncertainty in the D waves, which is believed to be under control.

\subsection{Energy Dependent Multipoles}
To parametrize the multipoles we have employed three different approaches
which we have fitted to the data:
(i) Empirical fit \cite{FBD09b,CD12,Hornidge2013} which depends on eight parameters
and  serves as a consistency test to single-energy multipole extraction; 
(ii) Heavy Baryon Chiral Perturbation Theory (HBCHPT) \cite{FB13} 
which depends on five parameters; 
and (iii) Relativistic Baryon Chiral Perturbation Theory (RBCHPT) \cite{BCHPT} 
which depends on five parameters.
In Fig. \ref{fig:experimental} we compare the obtained multipoles from these fits (dashed curves) 
to the extracted single-energy multipoles. The empirical fit provides 
a good description of both data and multipoles up to $E_\gamma=185$ MeV 
where the imaginary part of $M_{1+}$ ($\Delta$)
starts to provide a sizeable contribution. 
Both HBChPT and RBChPT provide a good description of data and multipoles up to 
$E_\gamma \approx 170$ MeV. Above that energy they fail to 
reproduce the differential cross section although still
provide very good results for the photon beam asymmetry.

Once the multipoles have been extracted, one can compare to predictions 
from the available literature.
In Fig. \ref{fig:experimental} we compare to four calculations
designed to work best in the resonance region: 
(i) Dispersive Effective Field Theory by Gasparyan and Lutz  (GL) \cite{GL}; 
(ii) Dubna-Mainz-Taipei (DMT) \cite{DMT};
(iii) Bonn-Gatchina (BoGa) \cite{BoGa};
(iv) and SN11 solution of SAID \cite{SAID_SN11}.
On the overall they provide a good description of the multipoles 
(solid lines in Fig. \ref{fig:experimental}), however, the deviations shown can
translate into large discrepancies with the actual data. 
The inclusion of the recent MAMI data in their analyses should improve the agreement
in this energy region together with a stringent constrain in the background contribution
to the resonance amplitudes \cite{Ronchen14}.

\section{Conclusions}
\begin{enumerate}
\item A2 and CB-TAPS Collaborations at MAMI have measured the differential cross section 
and the photon beam asymmetry in the near threshold region 
for the process of neutral pion photoproduction from the proton. 
The energy dependence of the photon beam asymmetry was obtained
 for the first time for this energy region.
\item The measurement of these observables has allowed to obtain 
the single energy S and P waves experimentally in a model independent way 
except for the uncertainty in the D waves, which is believed to be under control 
and impacts only the extraction of the S wave.
\item Several energy-dependent extractions of the multipoles have been performed: 
(i) Empirical \cite{Hornidge2013} 
-- that serves as a consistency test to single-energy multipole extraction,
provides a baseline to ponder on the quality of the other multipole extractions
and serves to predict other observables for future experiments; 
(ii) HBCHPT \cite{FB13}; 
(iii) RBCHPT \cite{BCHPT}. 
This last can be accessed through the MAID webpage \cite{ChiralMAID}. 
\item Current Chiral perturbation theory calculations only work up to 170 MeV of 
photon energy in the laboratory frame \cite{Hornidge2013,FB13,BCHPT}. 
Above this energy the theory calls for further improvement, 
such as the inclusion of the $\Delta$ \cite{DeltaBCHPT}. 
\item Lack of unitarity in the HBChPT amplitudes \cite{FBD09b} 
is not responsible for the disagreement
between theory and data \cite{FB13}.
\item Current theoretical calculations designed to work in the resonance region
\cite{GL,DMT,BoGa,SAID_SN11} provide a good description 
of the multipoles in the near threshold region. 
We expect that once these data are incorporated in the fits the agreement 
will improve and will help on achieving a better understanding 
of the background in such calculation \cite{Ronchen14}.
\item Data on the F and T asymmetries in the same energy region are 
currently under analysis \cite{FandTexperiment}.
\end{enumerate}

\begin{acknowledgement}
 
The author thanks A.M. Bernstein, D. Hornidge, and the
A2 and CB-TAPS Collaborations for providing the experimental data prior to publication,
and M. Hilt and L. Tiator for providing the RBCHPT calculation.
This research has been conducted with support 
of the Spanish Ministry of Science and Innovation grant
FIS2012-35316 and by CPAN, CSPD-2007-00042@Ingenio2010.  C.F.-R. is
supported by \textquotedblleft Juan de la Cierva\textquotedblright$\:$ 
program of the Spanish Ministry of Economy and Competiveness (Spain).
C.F-R. gratefully acknowledges the Organizers for their hospitality during the conference.
\end{acknowledgement}

\end{document}